\documentclass[preprint]{JHEP3} 

\JHEPspecialurl{http://jhep.sissa.it/JOURNAL/JHEP3.tar.gz}
\usepackage{graphicx}
\usepackage[square,comma,sort&compress]{natbib}

\newcommand\fverb{\setbox\fverbbox=\hbox\bgroup\verb}
\newcommand\fverbdo{\egroup\medskip\noindent%
			\fbox{\unhbox\fverbbox}\ }
\newcommand\fverbit{\egroup\item[\fbox{\unhbox\fverbbox}]}
\newbox\fverbbox

\newcommand{\qb}{{\bar q}}
\newcommand{\ub}{{\bar u}}

\newcommand{\nn}{\nonumber}
\newcommand{\ba}{\begin{equation}}
\newcommand{\ea}{\end{equation}}
\newcommand{\be}{\begin{eqnarray}}
\newcommand{\ee}{\end{eqnarray}}

\def\ib{{\bar\imath}}

\def\qb{{\bar q}}

\def\si{\sigma}

\def\A{{\cal A}}

\def\cD{{\cal D}}

\def\tree{{\rm tree}}

\def\oneloop{{1 \mbox{-} \rm loop}}

\title{Generalized unitarity at work: first NLO QCD results 
for hadronic {\boldmath $W+3$}~jet production} 
\author{R.~Keith~Ellis\\ Fermilab, Batavia, IL 60510, USA\\ 
Email: \email{ellis@fnal.gov}
}
\author{Kirill~Melnikov \\Department of Physics and Astronomy, Johns Hopkins 
University, Baltimore, MD 21218, USA\\
Email: \email{melnikov@pha.jhu.edu}
}
\author{Giulia Zanderighi \\Rudolf Peierls Centre for Theoretical Physics, 1 Keble Road, University of
   Oxford, UK\\
Email: \email{g.zanderighi1@physics.ox.ac.uk}
}

\received{\today} 		
\accepted{\today}		

\preprint{Fermilab-PUB-09-20-T\\
          OUTP-09-02-P} 

\abstract{We compute the leading color, next-to-leading order QCD
corrections to the dominant partonic channels for the production of a
$W$ boson in association with three jets at the Tevatron and the LHC.
This is the first application of generalized unitarity for realistic
one-loop calculations.  The method performs well in this non-trivial
test and offers great promise for the future.  }

\begin{document} 

\section{Introduction}

There are many multi-particle processes, knowledge of which through
next-to-leading order (NLO) in QCD would be very
desirable~\cite{Bern:2008ef}.  This statement, often repeated in the
context of the forthcoming experiments at the LHC is, in fact, true
even at the Tevatron. For example, the production rates for $W(Z)+n$
jets, with $n \le 4$ are well
measured~\cite{Aaltonen:2007ip,Aaltonen:2007cp} at the
Tevatron\footnote{Currently, for total cross-sections, errors range
from ten percent for $W+1$ jet to fifty percent for $W+4$ jets.  The
error on $W+3$ jets production cross-section is about twenty
percent~\cite{Aaltonen:2007ip,Aaltonen:2007cp}.} but next-to-leading
order QCD computations for such processes exist only for $n \le
2$~\cite{Campbell:2002tg}.

Of course, there are good reasons for that. The NLO QCD computations
for processes with a large number of external particles are difficult,
both analytically and numerically; a list of well-known problems can
be found in~\cite{Bern:2008ef}.  The need to overcome these
difficulties has made the computation of one-loop multi-leg scattering
amplitudes the focus of much research. In recent years, three main
suggestions for possible solutions have emerged as a result.
 
First, it was argued, and demonstrated by explicit computations, that
traditional methods, where one starts from Feynman diagrams and
proceeds through a Passarino-Veltman~\cite{Passarino:1978jh} style
reduction, can be optimized and made highly
efficient~\cite{Davydychev:1991va,Duplancic:2003tv,Giele:2004iy,Ellis:2005zh,denner,vanHameren:2005ed,Binoth:2008uq,Denner:2005es,Ellis:2006ss,Bredenstein:2008zb,Arnold:2008rz}.
Second, it was also shown that pure numerical approaches to NLO
computations are feasible~\cite{Soper:1998ye, Soper:1999xk,
Soper:2001hu, Nagy:2006xy, Lazopoulos:2007ix, Lazopoulos:2007bv,
Lazopoulos:2008de}.

The third idea
is the use of generalized unitarity where one starts
from on-shell tree-level scattering amplitudes and recycles them into
loops.  The idea of generalized unitarity was proposed in
Ref.~\cite{Bern:1994zx} more than ten years ago.  Important physical
results obtained using this method~\cite{Bern:1997sc} have
demonstrated both its potential and limitations.  The techniques of
applying generalized unitarity were significantly developed in recent
years thanks to important advances in
Refs.~\cite{britto,Britto:2004tx,opp,Ellis:2007br,Ellis:2008kd}.
These developments led to the design of two generalized
unitarity algorithms \cite{Berger:2008sj,Giele:2008ve}.
These methods are seminumerical in the sense that they 
depend on the complete analytic knowledge of the
relevant scalar integrals~\cite{Ellis:2007qk}.

The computational algorithm suggested in Ref.~\cite{Giele:2008ve} is
employed in this paper; we will refer to it as $D$-dimensional
generalized unitarity.  Note that this method was recently used to
obtain results not currently attainable with other methods, see e.g.
Refs.~\cite{Giele:2008bc,Ellis:2008ir,Ellis:2008qc}.  However, an
apparent weakness of generalized unitarity is that no 
result for a physical process has been obtained within this
framework.\footnote{We distinguish between generalized unitarity and
application of the algorithm of Ref.~\cite{opp} to Feynman diagrams.
The latter method was employed for the computation of NLO QCD
corrections to a relatively simple physical process $pp \to VVV$ in
\cite{bopp}.}  This should be contrasted with the traditional tensor
reduction approaches which never lost contact with phenomenology and
are being constantly refined to accommodate new challenges.

\begin{table}
\begin{center}
\begin{tabular}{|c|c|c|c|}
\hline
                                     & $W^{\pm}$, TeV & $W^+$, LHC & $W^-$, LHC \\
\hline
    $\sigma$ [pb], $\mu=40$~GeV      &  74.0 $\pm$   0.2  &   783.1 $\pm$  2.7 & 481.6 $\pm$ 1.4 \\
    $\sigma$ [pb], $\mu=80$~GeV      &  45.5 $\pm$   0.1  &   515.1 $\pm$  1.1 & 316.7 $\pm$ 0.7 \\
    $\sigma$ [pb], $\mu=160$~GeV     &  29.5 $\pm$   0.1  &   353.5 $\pm$  0.8 & 217.5 $\pm$ 0.5 \\
\hline
\end{tabular}
\end{center}
\caption{
The leading order total cross section for the production of a $W$ boson in association
with three jets including both two quark and four quark processes
vs. factorization and normalization scale.  The results are obtained
using the program MCFM. Cuts for the jets are $p_T>15$~GeV, $|\eta|<2$
at the Tevatron ($\sqrt{s} = 1.96$ TeV) and $p_T>50$~GeV, $|\eta|<3$
at the LHC ($\sqrt{s} = 14$ TeV). The CTEQ6L1 parton distributions
which have $\alpha_s(M_Z)=0.13$ are used.  The quoted errors are
statistical only.  }
\label{full_color_scale}
\end{table}

This is not a good situation for generalized unitarity which has to
live up to the claim of its advocates that it is a more powerful
method.  The only way to address this potential criticism is to
demonstrate the applicability of generalized unitarity in actual
calculations of direct phenomenological interest, preferably in
processes which are beyond the reach of traditional methods.  We have
chosen the production of a $W$ boson in association with three jets
for this purpose.  The reasons for our choice are as follows:

\begin{itemize} 

\item the calculation of NLO QCD corrections to this process is of
      direct
relevance since it is measured at the Tevatron~\cite{Aaltonen:2007ip,Aaltonen:2007cp}; it is not possible to 
use the leading order (LO) prediction for a serious comparison of
theoretical and experimental results because the LO cross section
varies by as much as a factor of two under reasonable changes in
renormalization and factorization
scales, see e.g. Table~\ref{full_color_scale};

\item measurements at the Tevatron have shown that for $W+n$ jets with $n=1$ 
and $2$, the data~\cite{Aaltonen:2007ip,Aaltonen:2007cp} is well
described by NLO QCD~\cite{Campbell:2002tg}; it is interesting to
verify this also for three and higher numbers of jets;

\item $W+3$ jet production 
is of interest for the LHC, being one of the backgrounds to
model-independent searches for new physics using the jets plus missing
energy signal;

\item the calculation of NLO QCD corrections to
$W+3$ jet production is highly non-trivial: there are 1583 Feynman
diagrams including a significant number of high-rank six-point
functions. There is no doubt that the computation of NLO QCD
corrections to this processes is a challenging task for traditional
diagrammatic approaches.

\end{itemize}

In our opinion these reasons make $W+3$ jet production an ideal
testing ground for the unitarity method and, recently, we made the
first step in that direction.  In Ref.~\cite{Ellis:2008qc} the three
current authors, together with Giele and Kunszt, applied the
generalized $D$-dimensional unitarity method to compute all one-loop
matrix elements needed for the NLO corrections to $W+3~{\rm jet}$
production including two-quark ($q \bar q ggg W$) and four-quark ($q
\bar q Q \bar Q W$) partonic processes~\cite{Ellis:2008qc}. Leading color 
two-quark amplitudes were also computed in Ref.~\cite{Berger:2008sz}.

With the virtual corrections in hand, two more steps are required to
arrive at physical predictions for $W+3~{\rm jet}$ production.  First,
the virtual corrections should be integrated over the relevant
phase-space.  Second, we need to consider processes with one
additional parton in the final state. When this parton becomes soft or
collinear to other partons in the process, the final state that
consists of four partons contributes to the final state with three
hard jets.

\begin{table}
\begin{center}
\begin{tabular}{|c|c|c|c|c|}
\hline
                          & $W$, $\sqrt{s}=1.96$~TeV,& $W^+$, $\sqrt{s}=14$~TeV, & $W^-$, $\sqrt{s}=14$~TeV,   \\
\hline
  $\sigma_{2q}^{\rm full}$ [pb]           & 29.63 $\pm$    0.04           &  356.99 $\pm$   0.77   &  216.35 $\pm$   0.40 \\
\hline
  $\sigma_{2q}^{\rm lc}$ [pb]           & 32.96 $\pm$    0.02             &
  377.08 $\pm$   0.79   &  229.89 $\pm$   0.42 \\
\hline
  $\sigma_{4q}^{\rm full}$ [pb]           &   16.13 $\pm$    0.03            &  147.60 $\pm$   0.38  &  94.91 $\pm$   0.19    \\
\hline
  $\sigma_{4q}^{\rm lc}$ [pb]           &  16.12 $\pm$    0.02         &  153.36 $\pm$   0.36  &   97.64 $\pm$   0.22   \\
\hline
\end{tabular}
\caption{Full and leading color 
cross sections for the production of a $W$ boson
and three jets for the two- (2q) and four-quark (4q) processes, 
at leading order.
Cuts, parton
distributions and scale choices as in Tab.~\protect\ref{full_color_scale}.
The renormalization and factorization scale is set equal to $80$~GeV.}
\label{total_sigmas}
\end{center}
\end{table}

\begin{table}
\begin{center}
\begin{tabular}{|c|c|c|c|c|}
\hline
                          & $W$, $\sqrt{s}=1.96$~TeV, $p \bar{p}$ & $W^+$, $\sqrt{s}=14$~TeV, $pp$ & $W^-$, $\sqrt{s}=14$~TeV, $pp$  \\
\hline
  $\sigma$ [pb]           & 32.96 $\pm$    0.02             &    377.08 $\pm$   0.79   &  229.89 $\pm$   0.42 \\
$gg$\, [\%]                & 2.60                               &  6.80                        &  11.16          \\
    $qg$\, [\%]                & 49.76                              &  37.90                       &  34.51          \\
    $gq$\, [\%]                & 2.35                               &  37.86                       &  34.42          \\
    $g\bar{q}$\, [\%]          & 19.89                              &  7.97                        &  9.33           \\
    $\bar{q}g$\, [\%]          & 3.63                               &  7.99                        &  9.38          \\
    $\bar{q}\bar{q}$\, [\%]    & 0.0                                &  0.0                         &  0.0          \\
    $qq$\, [\%]                & 0.0                                &  0.0                         &  0.0          \\
    $q\bar{q}$\, [\%]          & 21.52                              &  0.73                        &  0.60          \\
    $\bar{q}q$\, [\%]          & 0.26                               &  0.73                        &  0.60          \\
\hline
\end{tabular}
\caption{The cross section for the production of a $W$ and three jets at leading color
for the two quark processes and the percentages contributed by various incoming channels.
Cuts and parton distributions as in Table~\protect
\ref{full_color_scale}.
The renormalization and factorization scale is set equal to $80$~GeV.}
\label{leading_color_GFLAG}
\end{center}
\end{table}

In spite of all the technical improvements described above, the
computation of the matrix elements for $W+5$ partons at one-loop and
$W+6$ partons at tree-level and integrating them over the phase-space
are very challenging tasks.
For this reason it is useful to look for approximations, which help to
reduce the technical complexity of the problem and are justifiable
from a physics viewpoint. An obvious possibility is to consider the
large-$N_c$ approximation.

To check how well this approximation works, we study leading order
results for $W+3~{\rm jet}$ production.  A compilation of results
using the program MCFM~\cite{Campbell:2002tg} is given in
Tables~\ref{total_sigmas},\ref{leading_color_GFLAG}.
 It follows
from these Tables that, both at the Tevatron and the LHC, two-quark
processes dominate over four-quark processes.\footnote{We show numbers
for fixed renormalization and factorization scales, but the relative
decomposition into channels is largely scale-independent.}  At both
colliders two-quark processes provide about $70\%$ of the observed
cross-section, with four-quark processes being responsible for the
remaining $30\%$.  Also, the large-$N_c$ approximation turns out to be
good to about $10\%$ for both the Tevatron and the LHC.  We therefore
conclude that a useful first step towards computing NLO QCD
corrections to $W+3~{\rm jet}$ production cross-section is the
calculation of those corrections for partonic processes with {\it
only} two quarks in the initial and/or final state, in the large-$N_c$
approximation. Because the contribution of $gg$ channel is small both
at the Tevatron and the LHC, we may further limit ourselves to study
processes with at least a quark or an anti-quark in the initial state,
namely the six incoming channels $q\bar{q}, \bar{q}q, qg, \bar{q}g,
gq$, and $g\bar{q}$.  As we explain below, by working in the
large-$N_c$ approximation and by considering the two-quark channels
only, we can simplify the calculation significantly.

The remainder of the paper is organized as follows.  
 In Section \ref{sec:tree} the computation of real emission
corrections is described;  the integration of the $W+6$ parton 
matrix elements squared over the available  phase-space is 
discussed in Section~\ref{sec:int}.  
In Section~\ref{section:virt} the calculation 
of the virtual corrections performed in Ref.~\cite{Ellis:2008qc}
is reviewed.
In Section~\ref{sec:num} numerical results are presented. 
We conclude in Section~\ref{sec:conc}.

\section{Tree-level processes and subtraction terms }
\label{sec:tree}

In this Section we discuss the computation of the relevant tree-level
scattering amplitudes. We need these amplitudes to calculate the
production cross-section for $W+3~{\rm jets}$ at tree level as well as
the real emission correction to that process from the $W+4~{\rm
parton}$ final state. In what follows, we present matrix elements that
describe the production of a $W^+$ boson, but everything that we say
can be adapted to the case of $W^-$ production, after obvious
modifications.

The scattering amplitude for the process $0 \to \bar u +d + n~g+ W^+$
can be decomposed into color-ordered amplitudes according to the
equation
\begin{equation}
 {\cal A}_n^{\tree}(1_{\bar{u}},2_d,3_g,..,n_g)
 \ =\  g^{n-2} \sum_{\sigma\in S_{n-2}}
   (T^{a_{\sigma(3)}}..T^{a_{\sigma(n)}})_{i_2}^{~\ib_1}\
    A_n(1_{\bar{u}},2_d;\sigma(3)_g,..,\sigma(n)_g).
\label{qqngluons}
\end{equation}
In Eq.~(\ref{qqngluons}) $g$ is the strong coupling constant and
$S_{n-2}$ denotes the $(n-2)!$ permutations of the gluons.  Note that
neither the $W$ boson nor the electroweak couplings and CKM matrix
elements are displayed in Eq.~(\ref{qqngluons}).  We employ a
normalization of the color $SU(3)$ generators such that ${\rm Tr}(T^a
T^b)=\delta^{ab}$.

To calculate the production cross-section, we need to square the
matrix element in Eq.~(\ref{qqngluons}) and sum over the color and
spin degrees of freedom of the quarks and gluons. In the large-$N_c$
limit, individual color-ordered amplitudes do not interfere and we
obtain the scattering amplitude squared
\be
\sum _{\rm col, hel}
|{\cal A}_n^\tree(1_{\bar{u}},2_d,3,..,n)|^2 = 
\left ( g^2 \right )^{n-2} X_n
\sum \limits_{{\rm hel}, S_{n-2}}^{} 
|A_n(1_{\bar{u}},2_d;\sigma(3),..,\sigma(n))|^2,
\label{eq:largeN}
\ee
where $X_n = (N_c^2-1) N_c^{n-3}$. Note that we decided to keep 
some terms in the factor $X_n$ which are subleading in the 
large-$N_c$ limit.

To arrive at the production cross-section, we need to choose a
partonic initial state, square the scattering amplitude and integrate
it over the phase-space available for the final state particles. Each
choice of the initial state $ij$ leads to a particular number of
identical particles in the final state that we denote by $N_{ij}$.
When the integration over the phase-space available for final state
particles is performed, we have to divide the result by the symmetry
 factor $S_{ij} = N_{ij}!$. 
If we combine the symmetry properties of the phase-space with the fact that 
no interference terms are present in Eq.~(\ref{eq:largeN}), 
we can  reduce the number of scattering amplitudes that 
need to be calculated. 

For example, if the initial state is $u \bar d$, extreme
simplifications occur. In this case the final state is $W + (n-2)\;g$,
and the number of identical gluons is $N_{u \bar d} = n-2$, leading to
a symmetry factor $S_{u \bar d} = N_{u \bar d}\;! = (n-2)\;! =
S_{n-2}$.  We therefore write
\begin{eqnarray}
\sigma_{u \bar d}  && \sim 
\int \frac{{\rm d}\phi_{\rm glue} }{S_{\bar u d} }
\sum _{\rm col, hel} |{\cal A}_n^\tree|^2
=
\left(g^2\right)^{n-2}
\frac{X_n}{S_{\bar u d} } 
\int 
{\rm d}\phi_{\rm glue} 
\sum \limits_{{\rm hel},S_{n-2}}^{}
|A_n(1_{\bar{u}},2_d;\sigma(3),..,\sigma(n))|^2 
\nonumber \\
&& = \left ( g^2 \right )^{n-2} X_n \int {\rm d}\phi_{\rm glue}
\sum \limits_{\rm hel}^{} 
|A_n(1_{\bar{u}},2_d;3_{g_3}, 4_{g_4},..,.n_{g_n})|^2,
\label{Eq.sym}
\end{eqnarray}
where in the last step we used the symmetry of the $(n-2)$-gluon phase-space, 
${\rm d}\phi_{\rm glue}$, to argue 
that all color-ordered amplitudes give identical contributions to the 
cross-section. Since Eq.~(\ref{Eq.sym}) is the consequence of the fact
that gluons are identical particles, it holds true independently  
of cuts or other restrictions imposed on partons in the final state.

Other partonic channels can be simplified in a similar manner
although, typically, we gain less compared to 
the  $u \bar d$ initial state. For example, if we consider 
the initial state composed of a quark or anti-quark and a gluon,
there are $(n-3)$ identical gluons in the final state. Therefore, the
symmetry factor is $S_{qg} = (n-3)\,!$ and the 
 number of independent color-ordered
amplitudes that need to be considered is $S_{(n-2)}/S_{qg}  =
(n-2)$. For 3-parton final states $n=5$, so that there are three
independent amplitudes while for 4-parton final states, $n=6$ and the
number of independent amplitudes is four. When these numbers are compared 
to the $S_3 = 6$ and $S_4 = 24$ independent amplitudes that would be required 
if fixed ordering of the final state particles were discarded, 
we see that the improvement is substantial.

When the matrix element with $W+4$ partons in the final state is
integrated over the phase-space subject to the requirement that
three jets are observed, divergent results are obtained. These
divergences arise from phase-space regions where one of the four
partons in the final state becomes soft or two partons become
collinear to each other; eventually, they cancel against similar
divergences in the virtual corrections.  To achieve this cancellation
in practice, divergences in real emission contributions need to be
extracted. To make those divergences manifest, we use a subtraction
method~\cite{Ellis:1980wv} as formulated
 by Catani and Seymour~\cite{Catani:1996vz} who proposed
simple subtraction terms, which they called dipoles.  However, we need
to make minor modifications to the Catani-Seymour formalism because we
work with amplitudes where the ordering of identical particles in the
final state is fixed.

To find dipole subtraction terms consistent with fixed ordering of
the identical particles, it is convenient to follow the derivation in
Ref.~\cite{Catani:1996vz}. To this end, we calculate the soft limit of
the amplitude squared, partial fraction the eikonal factors to isolate
individual collinear limits, and extend the eikonal factors beyond the
soft limits, taking Ref.~\cite{Catani:1996vz} as an example.  The
dipoles that do not have soft singularities can be found by examining
collinear limits of the contributing amplitudes. Although identical
steps are required to find conventional dipoles, the difference
between dipole terms that employ ordered and full amplitudes can be
traced back to different soft limits and in the related necessity to
go beyond the soft limit in a different way.

We now illustrate the construction of the subtraction terms by
considering the $u \bar d$ initial state. Upon ordering the gluons in
the final state, the cross-section is determined by the square of a
single color-ordered amplitude summed over helicities.  In the actual
computation of the cross-section, this amplitude squared is multiplied
by the infrared-safe measurement function $F_J$ that depends on the
momenta of $n$ partons.  We therefore define
\be
\cD(2;3,4...n;1) = \sum \limits_{\rm hel} 
|A_n(1_{\bar u} ,2_{d},3,..,n)|^2 F_J(1,2,3,...n),
\ee
where labels $1$ and $2$  denote incoming particles. Any gluon in the 
final state can become soft. We 
calculate the soft limit of the amplitude squared and obtain
\begin{eqnarray}
\lim_{\rm soft} \cD(2;3,4,5,6;1) && =   
s(5,6,1)\;\cD(2;3,4,5;1) 
+s(4,5,6)\; \cD(2;3,4,6;1) \nonumber \\
&& + s(3,4,5)\; \cD(2;3,5,6;1) 
+s(2,3,4) \cD(2;4,5,6;1).
\label{eq2.5}
\end{eqnarray}
The eikonal factor in Eq.~(\ref{eq2.5})
\be
s(i,j,k) = \frac{ p_i p_k }{(p_i p_j)(p_j p_k)}
\ee
corresponds to the limit where momentum $p_j$ is soft.\footnote{We use 
a convention of treating all particles as if in the final state and 
as if all the momenta are outgoing. This allows us to write 
the soft limit in Eq.~(\ref{eq2.5}) in a symmetric way.}
Performing partial fractioning and extending eikonal factors 
beyond the 
soft limit,  we obtain the expression for the  subtraction 
term 
\begin{eqnarray}
&& \cD_{\rm sub}(2;3,4,5,6;1)= 
 \tilde D_{61,5}^{gq} \otimes \cD(2;3,4,\tilde 5;\widetilde{16}) 
+ \tilde D_{65,1}^{gg} \otimes  \cD(2;3,4,\widetilde{56};\tilde 1) 
\nonumber \\ 
&&
+ \tilde D_{54,6}^{gg} \otimes \cD(2;3,\widetilde{45},\tilde 6;1) 
+ \tilde D_{56,4}^{gg} \otimes \cD(2;3,\tilde 4,\widetilde{56};1) 
+ \tilde D_{43,5}^{gg} \otimes \cD(2;\widetilde{34},\tilde 5,6;1) 
\nonumber \\
&& 
+ \tilde D_{45,3}^{gg} \otimes \cD(2;\tilde 3,\widetilde{45},6;1) 
+ \tilde D_{34,2}^{gg} \otimes \cD(\tilde 2;\widetilde{34},5,6;1) 
+ \tilde D_{32,4}^{gq}  \otimes \cD(\widetilde{23};\tilde 4,5,6;1).
\label{aeq}
\end{eqnarray}
The notation $\widetilde{ij}$ and $\tilde j$ in Eq.~(\ref{aeq}) are
the standard notations, see Ref.~\cite{Catani:1996vz}.  The mapping
between momenta $p \to \tilde p$ that is required to evaluate the
right hand side in Eq.~(\ref{aeq}) is constructed in the same way as
in Ref.~\cite{Catani:1996vz}.  The dipole functions $\tilde
D_{ij,k}^{\rm {fl}_i,{\rm fl}_j} $ that we introduced in
Eq.~(\ref{aeq}) are closely related to the original Catani-Seymour
dipoles~\cite{Catani:1996vz} and we explain the exact correspondence
below. Before going into this, we point out that a modification of the
subtraction terms is required to remove the symmetry between the
emitter and emitted partons, inherent in final-final and final-initial
dipoles in the original formulation by Catani and
Seymour.\footnote{For initial-final and initial-initial dipoles this
symmetry is not present to begin with and we use standard expressions
for those dipoles.}  In our notation, the non-integrable singular
limit of the dipole $\tilde D_{ij,k}$ is associated with the soft
limit of parton $i$ whereas the soft limit of parton $j$ does not
introduce a non-integrable singularity.

To make things clear, we 
give examples of dipoles that we employ in the
present calculation.
For final-final dipoles, where both emitter and emitted partons are gluons, 
we use
\begin{eqnarray}
&& \tilde D_{ij,k}^{gg} \otimes \cD(2;..\widetilde {ij},\tilde k..;1) = 
\frac{1}{(p_i p_j) } 
\left [
-g_{\mu \nu} \left ( 
\frac{1}{1-z_{j} (1-y_{ij,k})}- 1 
\right )
\right. 
\nonumber \\
&& \left.
+ (1-\epsilon) \frac{l_\mu l_\nu }{2 p_i p_j }\right ]
A_5(2;..\widetilde {ij}_\mu, \tilde k..;1) 
A_5^*(2;..\widetilde {ij}_\nu, \tilde k..;1),
\label{eq:ffggdip}
\end{eqnarray}
where $\epsilon = (4-D)/2$ is the parameter of dimensional regularization, 
and  $y_{ij,k} = p_ip_j/(p_ip_j+p_ip_k+p_jp_k)$, 
$z_j = p_jp_k/(p_i p_k+p_j p_k)$ and 
$l = (1-z_j) p_i - z_j p_j$. The momenta of particles $\widetilde{ij}$ 
and $\tilde k$ are \cite{Catani:1996vz}
\be
p_{\widetilde{ij}} = p_i + p_j -\frac{y_{ij,k}}{1-y_{ij,k}} p_k,
\;\;\;
p_{\widetilde{k}} = \frac{p_k}{1-y_{ij,k}}.
\ee
In Eq.~(\ref{eq:ffggdip}) $A(..ij_\mu...)$ denotes the tree level
amplitude with the polarization vector for particle $ij$ removed.

For final-initial dipoles, where both emitter and emitted partons are gluons, 
we use
\begin{eqnarray}
&& \tilde D_{ij,a}^{gg} \otimes \cD(..\widetilde {ij},..\tilde a) = 
\frac{1}{p_i p_j x_{ij,a}} 
\left [
-g_{\mu \nu} \left ( 
\frac{1}{1-z_{j} +(1-x_{ij,a})}- 1 
\right )
\right. 
\nonumber \\
&& \left.
+ (1-\epsilon ) \frac{l_\mu l_\nu }{2 p_i p_j  }\right ]
A_5(..\widetilde {ij}_\mu,....\tilde a) 
A_5^*(...\widetilde {ij}_\nu,...\tilde a),
\end{eqnarray}
where $x_{ij,a} = 1 + p_i p_j/(p_ip_a+p_jp_a)$
and $z_j = p_j p_a/(p_i p_a + p_j p_a)$ and 
$l = (1-z_j) p_i - z_j p_j$. Note that the plus-sign between the 
first and the second term in the equation for $x_{ij,a}$ is the 
consequence of the all-outgoing momentum convention.

For initial-final dipoles, where emitter and emitted parton 
are quark and gluon respectively, we use 
\be
\!\!\! \tilde D_{ia,k}^{gq} \otimes \cD(\tilde k..\widetilde {ia}) \!= 
\!\frac{1}{2 (p_i p_a) x_{ia,k}} \!
\left [
\!\frac{2}{1-x_{ia,k} +u_i}- (1+\epsilon)
-x_{ia,k} (1-\epsilon)
\right ]\!
|A_5(\tilde k..\widetilde {ia})|^2,
\ee
where $x_{ia,k} = 1 + p_ip_k/(p_ip_a+p_kp_a)$, 
$u_i = p_i p_a/(p_i p_a+p_k p_a)$.

The subtraction terms for all other partonic channels are constructed
along similar lines. Since more orderings contribute to the amplitude
squared for partonic channels other than $u \bar d$ and $\bar d u$,
more dipoles need to be considered to account for all the singular
limits. Initial-initial dipoles appear for all channels except 
$u \bar d$ and $\bar d u$. Finally, let us note that the need to subtract
certain dipoles cannot be established from the soft limit of the
amplitude. In those case, the analysis of collinear singularities of
the amplitude squared is required to determine the dipoles that need to be
subtracted.

We also note that a simple but extremely useful modification of the
Catani-Seymour dipoles was suggested by Nagy~\cite{Nagy:2003tz}.  The
idea is to limit the subtraction to a small region of phase-space
available for final-state particles. To this end, final-final dipoles
are multiplied by $\theta(\alpha - y)$, final-initial dipoles by
$\theta (\alpha - (1-x))$, initial-final dipoles by $\theta ( \alpha -
u)$ and initial-initial dipoles by $\theta(\alpha - v)$ where $y,x,u$
and $v$ are standard variables used in \cite{Catani:1996vz}. This has
the advantage that the subtraction is not  performed if the kinematics
of four-parton final state is far away from the  singular limit, leading to
a considerable saving in computing time, because the matrix element
squared associated with the excluded dipole need not be computed.

As we have seen, the construction of the dipoles relevant for our
purposes is straightforward and requires only small modifications
compared to the original formalism of Catani and Seymour.  The next
step in the subtraction program -- the integration of the subtracted
terms over the unresolved phase-space -- is even more straightforward
since the integrals of our modified dipoles can be easily extracted
from the results quoted in Ref.~\cite{Catani:1996vz}.  To see this,
note that for initial-final and initial-initial dipoles, we do not
introduce any modifications relative to Ref.~\cite{Catani:1996vz}. We
do modify final-final and final-initial dipoles, but there is a simple
relationship between our dipoles $\tilde D$ and the ones in
Ref.~\cite{Catani:1996vz}, $D_{\rm CS}$. For example, for final-final
dipole, we can write
\be
 D_{\rm CS}(z,y) = \tilde D(z,y) + \tilde D(1-z,y).
\ee
To compute the integral of the subtraction term, we need to integrate 
$\tilde D(z,y)$ over $y$ and $z$
\be
\int \limits_{0}^{1} {\rm d} y\; {\rm d} z f(z,y) \tilde D(z,y), 
\ee
where $f(z,y)$ is the weight function. The important property of the
weight function is that it is symmetric with respect to $z \to 1-z$
transformation. Because of this symmetry property, we conclude that
\be
2 \int \limits_{0}^{1} {\rm d} y\; {\rm d} z f(z,y) \tilde D(z,y) 
= \int \limits_{0}^{1} {\rm d} y\; {\rm d} z f(z,y)  D_{\rm CS}(z,y). 
\ee
Since the integral that appears on the right hand side of that equation 
is computed in \cite{Catani:1996vz}, $\int \tilde D(z,y)$ can 
be easily extracted. A similar reasoning can be used to obtain 
integrals of the final-initial dipoles.

\section{Integration over the phase-space} 
\label{sec:int}

The next issue to be discussed is the integration of the difference
between the matrix element squared and the subtraction term over the
entire phase-space allowed by the external cuts.  We use VEGAS 
\cite{Lepage:1977sw}
to adapt the integration grid automatically but we still need to generate
the parton kinematics carefully to ensure efficient sampling.

In addition, when trying to integrate over the phase-space, we face a
difficulty inherent in any subtraction method. To illustrate the
issue, consider a matrix element squared that requires a subtraction
of a particular final-final dipole to make it integrable.  The dipole
is described by two standard variables $y$ and $z$. The matrix element
squared has non-integrable singularities at $y=0$, $z=0$ and at $y=0$,
$z=1$; these singularities are removed by subtraction.  The difference
between the matrix element and the subtraction term is still singular,
but these singularities are integrable. We assume that the difference
scales as $1/\sqrt{y}$ and as $1/\sqrt{z}$ or $1/\sqrt{1-z}$.
Although these are integrable singularities, in order to have the
standard estimate of the integration error when using Monte-Carlo
integration techniques, it is mandatory to change integration
variables to absorb the square-root singularities into the measure.

Since we have a left-over square-root singularity for every dipole
that we need to subtract from the matrix element squared, we need to
do a large number of variable transformations if we want to absorb all
the singularities.  Note also that different dipoles need to be
subtracted for different initial states. This means that the required
changes of variables can not be done globally and, instead, we have to
adopt a multichannel integration technique.  To this end, for each 
partonic channel,  we
\begin{itemize}
\item randomly pick a dipole that contributes to a chosen 
channel -- all dipoles are given equal weights;
\item generate the phase-space for the chosen 
dipole in such a way that the  square-root singularity 
can be  absorbed into the integration measure. 
\end{itemize}

More specifically, suppose we are interested in the contribution of a
particular partonic channel to the production cross-section for which
$N_d$ dipoles are required to make it finite. We need to compute
\be
I = \int {\rm d} \phi_4 \left ( |A|^2 - |A|^2_{\rm subtr} \right ), 
\ee
where ${\rm d} \phi_4$ denotes the phase-space with four partons 
in the final state, $|A|^2$ is the matrix element squared 
and $|A|^2_{\rm subtr}$ is the subtraction term. We rewrite the 
integral as  
\be
I = \sum_{n=1}^{N_d} \alpha_n I_{n}, 
\ee
where $\alpha_n$ are some constants and
\be
I_{n} = \int  \frac{{\rm d} \phi_4}{J_n} 
\frac{ \left ( |A|^2 - |A|^2_{\rm subtr} \right )}{\sum \limits_{m} \alpha_m {J_m}^{-1}}.
\label{eq3.3}
\ee
In Eq.~(\ref{eq3.3}) $J_n$ are Jacobian factors. In the actual 
calculation, we choose the coefficients $\alpha_n$ to be equal 
although, in principle, it is possible to optimize their choice iteratively.

We associate each $I_n$ with a particular dipole contribution that
needs to be subtracted from the matrix element squared. Suppose that
$n$ is a final-final dipole.  We start by generating the phase-space
for {\it three massless} partons and the $W$ boson, assuming that the
kinematics of the massless parton is characterized by a ${\rm d}
p_\perp/p_\perp$ distribution (for transverse momentum $p_T$ above the
jet $p_T$ cut) and by a uniform distribution in
rapidity.\footnote{This is the standard procedure in MCFM.}  Then, we
use the fact that the four-parton phase-space factorizes into the
product of the three-parton phase-space and the dipole phase-space
that is completely specified by three additional variables. We denote
the momenta of the three-parton final state as $\tilde p$, and momenta
of the four-parton final state as $p$. Additional variables needed to
describe the kinematics of the $3 \to 4$ splitting are $y,z$ and
$\phi$. We therefore write
\be
{\rm d} \phi_{4} (p) = {\rm d} \phi_{3}(\tilde p_{ij},\tilde p_k) 
{\rm d} p_i(\tilde p_{ij},p_k),
\ee
and 
\be 
{\rm d} p_i (\tilde p_{ij},p_k) = 
\frac{\tilde p_{ij} \tilde p_{k}}{8 \pi^2}
\; \frac{{\rm d} \phi}{2\pi} \; {\rm d} z_i \; {\rm d}y_{ij,k} \; (1-y_{ij,k}).
\ee

Having  generated the momenta of three-parton final state $\tilde p$, 
we use them to construct the momenta $p$ 
according to the following formulae
\begin{eqnarray}
&& p_i = z_i \tilde p_{ij} + y_{ij,k} z_j \tilde p_k + k_\perp,\nonumber \\
&& p_j = z_j  \tilde p_{ij} + y_{ij,k} z_i \tilde p_k - k_\perp,  \\
&& p_k = \left (1-y_{ij,k}\right ) \tilde p_k, \nonumber \\
&& p_{m \ne i,j,k} = \tilde p_m. \nonumber 
\label{ffdip}
\end{eqnarray}
The transverse momentum reads 
$k_\perp^\mu = |k_\perp| \left ( \cos \phi \; v_1^\mu +   
\sin \phi \; v_2^\mu \right )$ and 
$|k_\perp| = \sqrt{y z(1-z) 2 \tilde p_{ij} \tilde p_k }$.
The two auxiliary vectors $v_{1,2}$  are such that 
$v_{1,2}^2 = -1$,  $v_{1} v_{2} = 0$, 
$v_{1,2} \tilde p_{ij} = 0$, 
$v_{1,2} \tilde p_{k} = 0$.

The choice of the Jacobian factor $J_n$ allows us to absorb the
square-root singularities; for final/final dipoles we use the form
suggested in Ref.~\cite{Seymour:2008mu}:
\be
J_n^{-1} = \frac{\sqrt{z_i} + \sqrt{z_j}}{\sqrt{y_{ij,k}}\sqrt{z_i z_j}}.
\label{eq3.8}
\ee

Note that this Jacobian can be written in terms of the four parton momenta $p$,
since, as follows from Eq.~(\ref{ffdip}), we 
can express $z_i$, $z_j$ and $y_{ij,k}$ in terms of these momenta,
\be
z_i  = \frac{p_i p_k}{p_i p_k + p_j p_k},\;\;\;
z_j = 1-z_i,\;\;\; y_{ij,k} = \frac{p_i p_j}{p_i p_k + p_j p_k+p_i p_j}. 
\label{eq3.9}
\ee
This remark is important since, as follows from Eq.~(\ref{eq3.3}),
each integral $I_n$ requires the knowledge of all Jacobians $J_m$
including the ones with $m \ne n$. This, however, is not a problem
because all those Jacobians are {\it uniquely} expressed through
momenta $p$ in the spirit of Eqs.~(\ref{eq3.8}) and~(\ref{eq3.9}).

The Jacobian $J_n$ can be absorbed into the measure by a simple 
change of variables
\be 
\frac{{\rm d} p_i (\tilde p_{ij},p_k)}{J_n} = 
\frac{\tilde p_{ij} \tilde p_{k}}{\pi^2}
\; \frac{{\rm d} \phi}{2\pi} \; {\rm d} \mu_{ij,k} \; {\rm d} \xi_{i}
\; (1-y_{ij,k}) ,
\ee
where
\be
\bar{\phi}=\frac{\phi}{2 \pi},\;\;\;\;y_{ij,k} = \mu_{ij,k}^2,\;\;\;\;
z_i = \frac{1}{2} \left ( 1 - (1-2\xi_i) \sqrt{1+4\xi_i(1-\xi_i)} \right ),
\ee
with $ 0< \mu_{ij,k} < 1$ and $0 < \xi_i < 1$. So, given random
numbers for $\mu,\xi,$ and $\bar{\phi}$ and the momenta of
three-parton final state $\tilde p$, we can calculate $z$ and $y$, the
phase-space weight and the corresponding momenta $p$ of the
four-parton final state. We then calculate the matrix element squared
as discussed in the previous Section and derive a particular
contribution to the differential cross-section for $W+{\rm 3}~{\rm
jet}$ production.

Another case which requires comment involves dipoles where initial
particles are present. We will consider the final-initial dipoles as
an example.  The general strategy is very similar to what has already
been discussed.  We start by generating momenta for the three-parton
final state and the $W$-boson and three random numbers $z, x$ and
$\phi$. We denote the momenta of one of the partons in the final state by
$\tilde p_{ij}$ and the momentum of the spectator in the initial state
by $\tilde p_a$. We then generate the additional parton momentum
according to the following formulae
\begin{eqnarray}
&& p_i = z \tilde p_{ij} + (1-z) \frac{1-x}{x} \tilde p_a 
+ k_\perp, \nonumber \\
&& p_j = (1-z) \tilde p_{ij} + z \frac{1-x}{x} \tilde p_a 
- k_\perp, \nonumber \\
&& p_a = \frac{\tilde p_a }{x},  \\
&& p_{m \ne i,a,k}= \tilde p_m ,  
\end{eqnarray}
where 
 $k_\perp^\mu = |k_\perp| \left ( \cos \phi \; v_1^\mu +   
\sin \phi \; v_2^\mu \right )$, $v_{1,2}^2 = -1$, 
$v_{1,2} \tilde p_{ij} = 0$, 
$v_{1,2} \tilde p_{a} = 0$ and 
$|k_\perp| = \sqrt{2 \tilde p_{ij} p_a z (1-z) (1-x)}$.

For the final-initial dipole, we employ the Jacobian 
\be
J_n^{-1} = \frac{\sqrt{z} + \sqrt{1-z}}{\sqrt{1-x}\sqrt{z (1-z)}}.
\ee
To absorb this Jacobian into the integration measure, we make a change
of variables along the lines discussed in connection with final-final
dipoles.  Furthermore, we have to make sure that the new momenta of
the initial parton $p_a$ does not exceed the momentum of the proton
since, a priori, the variable $x$ can assume any value between $0$ and
$1$. If the momentum of the initial state parton exceeds the momentum
of the proton, the corresponding event is rejected.  We deal with
initial-final and initial-initial dipoles along similar lines.

\section{Virtual corrections}
\label{section:virt}

A detailed description of the calculation of all one-loop amplitudes
needed for the NLO correction to $W$+3\,jets at hadron colliders is
given in Ref.~\cite{Ellis:2008qc}. Here we recall the elements of the
discussion needed for the purpose of this paper.

At one-loop, using the color basis of Ref.~\cite{DelDuca:1999rs} and
neglecting contributions from closed fermion loops, the color
decomposition for the process $0 \to \bar u +d + n~g+ W^+$ can be
written in terms of left primitive amplitudes~\cite{Bern:1994fz} as
\begin{eqnarray}
\label{qqngluonsoneloop}
\A_{n}^\oneloop(1_\ub,2_d,3_g,\ldots,n_g)
&=& g ^n  \biggl[
\sum_{p=2}^n \sum_{\si \in S_{n-2}} 
(T^{x_2} T^{a_{\si_3}} \cdots T^{a_{\si_p}}
 T^{x_1})^{~\bar{i}_1}_{i_2}
(F^{a_{\si_{p+1}}} \cdots F^{a_{\si_n}})_{x_1 x_2} \nn\\
&& \hskip -0.3 cm
\times (-1)^n A_n^{L}(1_\ub,{\si(p)}_g,\ldots,{\si(3)}_g,
2_d,{\si(n)}_g,\ldots,{\si(p+1)}_g)
\biggr]\!.
\end{eqnarray}
In Eq.~(\ref{qqngluonsoneloop}) for $p=2$ the factor ${(T\cdots
T)_{i_2}}^{\bar i_1}$ becomes ${(T^{x_2}T^{x_1})_{i_2}}^{\bar i_1}$
and for $p=n$ the factor $(F\cdots F)_{x_1x_2}$ becomes
$\delta_{x_1x_2}$.  As before neither the $W$ boson nor the
electroweak couplings and CKM matrix elements are displayed in
Eq.~(\ref{qqngluonsoneloop}).  In the leading color approximation we
retain only the $p=2$ term and Eq.~(\ref{qqngluonsoneloop}) simplifies
to
\begin{eqnarray}
\A_{n}^\oneloop(1_\ub,2_d,3_g,\ldots,n_g)
&=&  g ^n \biggl[
\sum_{\si \in S_{n-2}} 
(T^{x_2} T^{x_1})^{~\bar{i}_1}_{i_2}
(F^{a_{\si_{3}}} \cdots F^{a_{\si_n}})_{x_1 x_2} \nn\\
&& \hskip 0.5 cm
\times (-1)^n A_n^{L}(1_\ub,2_d,{\si(n)}_g,\ldots,{\si(3)}_g)
\biggr]\,.
\end{eqnarray}
The matrix element squared is then given by
\begin{eqnarray}
\label{eq:virtsquare}
&&2 \sum_{\rm col, hel} |\A_{n}^\oneloop(1_\ub,2_d,3_g,\ldots,n_g)
{\cal A}_n^{\tree, *}(1_{\ub},2_d,3_g,..,n_g) | \\
&&= 2\, (N_c^2-1)\,N_c^{n-2} \left(g^2\right)^{n-1}
\sum_{\rm hel, S_{n-2}}
|A_{n}^L(1_\ub,2_d,3_g,\ldots,n_g)
{A}_n^{*}(1_{\ub},2_d,3_g,..,n_g) |\nonumber\,,
\end{eqnarray}
where again we choose to keep some  subleading color terms. We can now 
take advantage of the symmetry of the phase-space and fix the ordering 
of identical particles in the final state. The procedure is described  
in detail in Section~\ref{sec:tree}. 

\section{Results}
\label{sec:num}

In this Section we present the results of our calculation of NLO QCD
corrections to $W+3~{\rm jet}$ production.  We begin by describing
computational aspects of the problem.  The total time needed for the
calculation of virtual corrections is determined by how many one-loop
primitive amplitudes must be evaluated and how computationally
expensive they are.  At leading color, we only need to calculate the
fastest primitive amplitude, for which about 50~ms are required for a
given momentum/helicity configuration.\footnote{All numbers are given
for a computer with 2.33~GHz Pentium Xeon processor and {\sf Intel}
fortran compiler.}

To compute the hadronic cross-section, we sum over six partonic
channels.  As explained in previous Sections, we fix the ordering of
gluons in the final state.  Then, for partonic channels with quark and
anti-quark in the initial state, we evaluate the virtual primitive
amplitudes $2^3 = 8$ times, since we sum over two helicities of each
of the three gluons. For the $qg, gq, \qb g, g\qb$ channels we
consider three different orderings of the final-state fermion,
relative to the gluons. As the result, we evaluate $24 = 3 \times
2^{3}$ primitive amplitudes per partonic channel with a single fermion
in the initial state.  Therefore, for the computation described in
this paper, we need to compute the leading color primitive amplitude
112 times for each phase-space point.  This translates into a total of
5.6 seconds per phase-space point.

This time is too large to allow us to compute the virtual corrections
on a dynamical, self-adapting grid.  We therefore adopt the following
strategy for computing virtual corrections.  First, we compute the
tree level cross-section with a large number of points, $2
\times 10^7$, and establish an integration grid.\footnote{There are
other ways to establish the grid. For instance, we can omit the
computationally expensive parts of the virtual amplitudes and keep
only logarithms of kinematic invariants that come from $1/\epsilon$
poles. We have checked that changing the strategy for establishing the
grid has no bearing on the final result.} Once the grid is fixed, we
compute virtual corrections by running three different jobs each with
$10^5$ evaluations using different seeds to start VEGAS off. We then
average the results of these three evaluations.  With this procedure,
Monte Carlo errors for virtual corrections are around $0.7-1$\%.

For real corrections, we need about $10~{\rm msec}$ per phase-space
point to compute matrix element squared and the subtraction
terms\footnote{This number depends on the value of the parameter
$\alpha$ that determines how often subtraction terms need to be
calculated~\cite{Nagy:2003tz}. The quoted value corresponds to $\alpha
= 0.01$.}.  Since computation of the real emission correction is
inexpensive, we calculate these corrections following the standard
MCFM procedure of first doing a pre-conditioning run and then a final
run. We used 10 times $4\cdot 10^6$ points for the pre-conditioning
run and five times $8\cdot 10^6$ points for the final run.
With this number of events, the  Monte Carlo integration errors
for the real (subtracted) contribution are around $0.4-0.7$\%.

As far as final results are concerned, errors coming from virtual 
and real corrections are comparable   and 
the total Monte Carlo integration error is in the range 
of $1-4$\%; this is better  than the theoretical
uncertainty of the results estimated with a standard renormalization
and factorization scale variation.

\vspace*{0.5cm} We are now in position to describe the numerical
results of the computation.  However, before we enter into this
discussion, we remind the reader that our results are approximate for
the following reasons:

\begin{itemize} 

\item  we employ the large-$N_c$ approximation to compute the scattering 
amplitudes;  using the leading order cross-section as a guide, we estimate 
that this approximation is accurate to about $10$ percent;

\item  we include only the two-quark processes   $q \bar q ggg W$ and 
ignore the four-quark processes $q \bar q Q \bar Q g W$. Even within
the two-quark processes, we do not consider the partonic channel with
two gluons in the initial state.  For the leading order cross-section,
we find that the four-quark processes increase the cross-section by
thirty percent so that omitting them gives results accurate to about
thirty percent.

\end{itemize}

Because of these approximations, we warn the reader that absolute
results for cross-sections and distributions that we report below
should be used with caution. We believe, however, that ratios of NLO
and LO results for various observables are less sensitive to these
omissions.

\begin{figure}[t]
\begin{center}
\includegraphics[angle=0,scale=0.9]{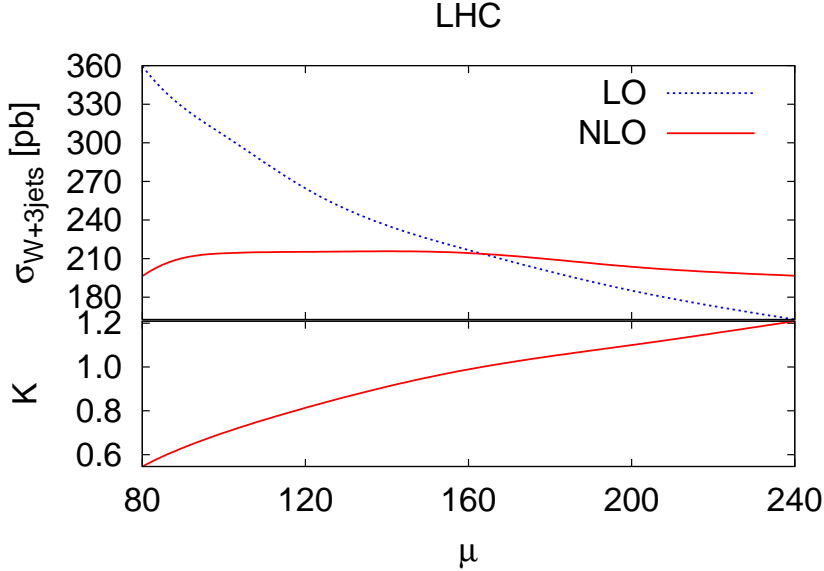}
\caption{
Inclusive $W^++3$ jet 
cross-section at the LHC and the $K$-factor defined 
as $K = \sigma_{\rm NLO}/\sigma_{\rm LO}$ 
as a function of the renormalization and factorization
 scales. 
Jets are defined with $k_T$ algorithm with $R= 0.7$ and $p_T>50$~GeV. 
Jet rapidities  satisfy  $|\eta|<3$.  The LO and NLO 
cross-sections are computed with CTEQ6L1 and CTEQ6M parton 
distributions, respectively.
}
\label{fig1}
\end{center}
\end{figure}

The numerical results for $W+3$ jet production at NLO are obtained
using the CTEQ6m parton distributions~\cite{Pumplin:2002vw} which have
a value of $\alpha_S(M_z)=0.118$. The evolution of the coupling
constant is performed using the two-loop beta function
\begin{equation}
\beta(\alpha_s) = -b \alpha_S^2 (1+ b^\prime \alpha_S),\quad b
=\frac{33-2 n_f}{12 \pi},\quad b' =\frac{153-19n_f}{2 \pi (33-2 n_f)}\,,
\end{equation}
where, in the spirit of the large-$N_c$ approximation, we set the
number of light flavors $n_f$ equal to zero.  The $k_T$ jet algorithm
with $R=\sqrt{\Delta\phi^2 + \Delta \eta^2} = 0.7$ 
and $p_T > 15~{\rm GeV}\; (p_T> 50~{\rm GeV})$ 
is used to define jet
cross sections at the Tevatron and the LHC, respectively. We employ
default MCFM choice for electroweak parameters and the CKM matrix
elements; they can be found in Ref.~\cite{Campbell:2002tg}.

\begin{figure}[t]
\begin{center}
\includegraphics[angle=0,scale=0.9]{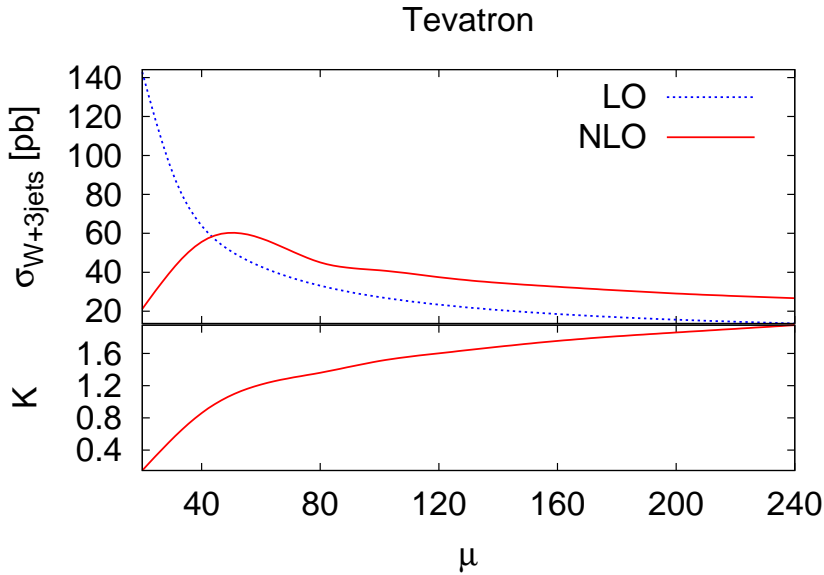}
\caption{The inclusive $W+3$ jet 
cross-section at the Tevatron and the $K$-factor defined as $K =
\sigma_{\rm NLO}/\sigma_{\rm LO}$ as a function of the renormalization
and factorization scales $\mu$.  Jets are defined with $k_T$ algorithm
with $R= 0.7$ and $p_T>15$~GeV.  Jet rapidities satisfy $|\eta|<2$.
The LO and NLO cross-sections are computed with CTEQ6L1 and CTEQ6M
parton distributions, respectively.}
\label{fig2}
\end{center}
\end{figure}

In Figs.~\ref{fig1},\ref{fig2} we present total cross-sections and
$K$-factors, defined as $K = \sigma_{\rm NLO}/\sigma_{\rm LO}$, for
$W+3$ jet production at the LHC and the Tevatron as a function of the
factorization and the renormalization scales which we set equal to
each other $\mu_R = \mu_F = \mu$.  At the LHC, the NLO cross-section
shows remarkable independence of the scale $\mu$, unlike the LO
result.  The equality of LO and NLO cross-sections occurs at $\mu_0
\approx 160~{\rm GeV}$. Because the dependence of the LO cross-section
on the unphysical scale $\mu$ is strong, the NLO corrections are
typically large. For example, choosing $\mu = m_W$ to compute the LO
cross-section for $W+3$ jet production at the LHC, leads to NLO QCD
corrections of the order of $-50\%$.

For the Tevatron, the situation is different. First, the dependence of
the NLO cross-section on the renormalization and factorization scales
is sizeable although it is significantly reduced compared to the
leading order cross-section. In addition, as follows from
Fig.~\ref{fig2} the equality of leading and next-to-leading order
cross-sections occurs at a scale $\mu_0 \approx 50~{\rm GeV}$ which is
much smaller than the LHC case discussed above.  This is not
unexpected since both the center of mass energy and the $p_\perp$ cut
for jets is smaller at the Tevatron which leads to a much softer
spectrum of jets compared to the LHC case.

It is interesting to note that gross features of NLO QCD corrections
to $W+3$ jet production, such as scales at which leading and
next-to-leading order cross-sections coincide, are very
similar to what was observed in NLO QCD computation of $W+2~{\rm
jets}$ \cite{Campbell:2002tg, Campbell:2003hd}. 
What differs between two and three jet production is the price one pays 
for making an infelicitous choice of
scale in the LO result.  Because the dependence on $\mu$ of
$\sigma_{W+3~{\rm jet}}^{\rm LO}$ is stronger than of
$\sigma_{W+2~{\rm jet}}^{\rm LO}$, $K$-factors for $W+3$ jet decrease
or increase stronger when one moves away from $\mu = \mu_0$.

\begin{figure}[t]
\begin{center}
\includegraphics[angle=0,scale=0.9]{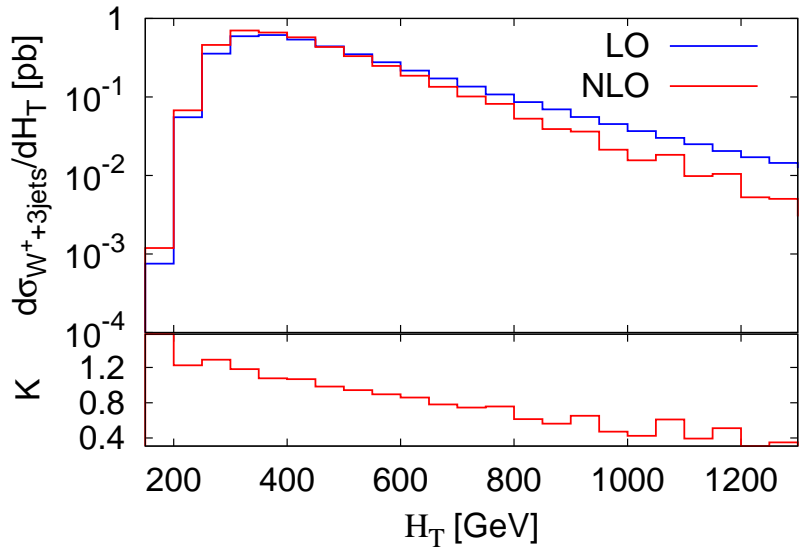}
\caption{
The distribution of the transverse energy 
$H_T = \sum \limits_{j}^{} E_{\perp,j} +E^{\rm miss}_{\perp}
+E_{\perp}^{e}$  and the $K$ factor defined as 
$K = ({\rm d}\sigma^{\rm NLO}/{\rm d}H_T)\;/
 ( {\rm d}\sigma^{\rm LO}/{\rm d}H_T )$
in $W^++3~{\rm jet}$ inclusive 
production at the LHC at leading and next-to-leading 
order. Renormalization and factorization scales are set 
to  $160$ GeV.}
\label{fig3}
\end{center}
\end{figure}

Finally, we present selected results for differential distributions at
the LHC. We choose the renormalization and factorization scales to be
$160~{\rm GeV}$ since this minimizes the inclusive $K$-factor. In
Fig.~\ref{fig3} we plot the distribution in the variable $H_T$ defined
as the sum of transverse energies of jets, the missing transverse
energy and the transverse energy of the lepton $\displaystyle H_T = \sum
\limits_{j}^{} E_{\perp,j} +E^{\rm miss}_{\perp} +E_{\perp}^{e}$.  The
variable $H_T$ measures the overall hardness of a particular event and
can be employed in model-independent searches for New Physics.  As
illustrated Fig.~\ref{fig3}, the $H_T$-distribution becomes softer at
NLO, at least in comparison to the leading order result calculated 
at fixed scale $\mu=160$~GeV.

\begin{figure}[t]
\begin{center}
\includegraphics[angle=0,scale=0.9]{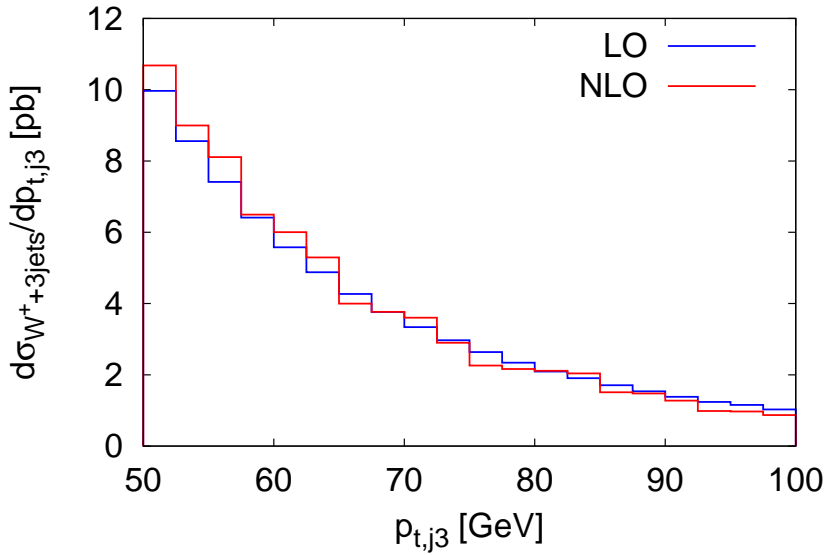}
\caption{
The transverse momentum distribution of the third hardest jet 
in the inclusive $W+3$ jet production at the LHC. 
The renormalization and factorization scales are set to $160$ GeV.}
\label{fig4}
\end{center}
\end{figure}

In Fig.~\ref{fig4} we present the transverse momentum distribution of
the third hardest jet in the inclusive production of $W+3$ jets at the
LHC.  In the range of $p_\perp$ shown in the plot, 
the shapes of $p_\perp$ distributions at leading 
and next-to-leading order are nearly identical.

\section{Conclusions} 
\label{sec:conc}
In this paper, we apply the method of generalized $D$-dimensional
unitarity~\cite{Giele:2008ve} to compute NLO QCD corrections to $W+3$
jet production at the Tevatron and the LHC.  There are two reasons
that make this result an important benchmark in the field of one-loop
QCD computations for hadron collider physics. First, this is the only
application of the idea of generalized unitarity in a fully realistic
one-loop computation that goes beyond calculation of one-loop helicity
amplitudes at a fixed point in phase-space. Second, our result
 is one of the very few  computations of one-loop corrections 
to six-parton processes at hadron colliders -- the current 
research frontier in  NLO QCD.  It is remarkable that 
the method achieved this benchmark without a problem; this assures us that 
generalized unitarity is a practical computation method 
that can be applied to other, perhaps even more complicated, processes.

Looking into the near future, we expect to refine our computation in
two ways. First, we expect to include the four-quark partonic channels
in the large-$N_c$ approximation; this is an important step for
realistic phenomenology.  Further down the road, we may want to extend
the computation beyond the leading color approximation.  Estimating
the increase in computer running time required to go beyond the large
$N_c$~limit, we find that about two minutes per point will be needed
to compute virtual corrections to the matrix element squared.  At face
value, this is feasible, but computationally expensive. However,
one can imagine various improvements, including Monte Carlo sampling
over helicities and colors, that should lead to an appreciable
improvement in the speed of the program.

Finally, we would like to say a few words about phenomenology. Since
we did not consider the four-quark channels in this paper, we decided
not to pursue very detailed phenomenological studies.  However, the
numerical results that we {\it do} report are instructive since they
give an idea about potential significance of NLO QCD effects in $W+3$
jet production at the Tevatron and the LHC.  Our computation shows
that NLO QCD effects are large and can reach $\pm 50 \%$, if
unfortunate, but not unreasonable, choices of the renormalization and
factorizations scales are made in a computation based on leading order
matrix elements.  Note that the probability that an unfortunate scale choice 
is made increases for a larger number of jets
since the production cross-section at LO becomes a steeper function of
the renormalization and factorization scales. The only way to cure
this problem is by computing NLO QCD corrections. For processes like
$pp \to W+4$ jets or $pp \to t \bar t + 2j$ this will be complicated 
no matter what method is used, 
but we believe that generalized unitarity will be up to the task.

\newpage

\section*{Acknowledgments} 
We are grateful to W.~Giele, Z.~Kunszt and G.~Salam for useful discussions.
The research of K.M. is supported by the startup package provided 
by Johns Hopkins University. G.Z. is supported by the 
British Science and Technology
Facilities Council.  Fermilab is operated by 
Fermi Research Alliance, LLC under Contract
No. DEAC02- 07CH11359 with the United States Department of
Energy.

\end{document}